%% file: ms.tex
\documentclass[12pt,preprint]{aastex}

\shorttitle{LSR1610}
\shortauthors{Cushing et al.}

\begin{document}


\title{The Schizophrenic Spectrum of LSR 1610$-$0040:  a Peculiar M Dwarf/Subdwarf}


\author{Michael C. Cushing\altaffilmark{1,2}}
\affil{Steward Observatory, University of Arizona, 933 North Cherry Avenue, Tucson, AZ 85721}
\email{mcushing@as.arizona.edu}

\and

\author{William D. Vacca\altaffilmark{1}}
\affil{SOFIA-USRA, NASA Ames Research Center MS N211-3, Moffett Field, CA 94035} \email{wvacca@mail.arc.nasa.gov}

\altaffiltext{1}{Visiting Astronomer at the Infrared Telescope Facility, which is operated by the University of Hawai`i under cooperative Agreement no. NCC 5-538 with the National Aeronautics and Space Administration, Office of Space Science, Planetary Astronomy Program.}
\altaffiltext{2}{Spitzer Fellow}

\begin{abstract}

We present a moderate resolution ($R$ $\equiv$ $\lambda/\Delta\lambda$ $\approx$ 2000), 0.8$-$4.1 $\mu$m spectrum of LSR 1610$-$0040, a high proper motion star classified as an early-type L subdwarf by L{\'e}pine and collaborators based on its red-optical spectrum.  The near-infrared spectrum of LSR 1610$-$0040 does not fit into the (tentative) M/L subdwarf sequence but rather exhibits a mix of characteristics found in the spectra of both M dwarfs and M subdwarfs.  In particular, the near-infrared spectrum exhibits a \ion{Na}{1} doublet and CO overtone bandheads in the $K$ band, and \ion{Al}{1} and \ion{K}{1} lines and an FeH bandhead in the $H$ band, all of which have strengths more typical of field M dwarfs.  Furthermore the spectrum of Gl 406 (M6 V) provides a reasonably good match to the 0.6$-$4.1 $\mu$m spectral energy distribution of LSR 1610.  Nevertheless the near-infrared spectrum of LSR 1610 also exhibits features common to the spectra of M subdwarfs including a strong \ion{Ti}{1} multiplet centered at $\sim$0.97 $\mu$m, a weak VO band at $\sim$1.06 $\mu$m, and possible collision-induced H$_2$ absorption in the $H$ and $K$ bands.  We discuss a number of possible explanations for the appearance of the red-optical and near-infrared spectrum of LSR 1610$-$0040.  Although we are unable to definitively classify LSR 1610$-$0040, the preponderance of evidence suggests that it is a mildly metal-poor M dwarf.  Finally, we tentatively identify a new band of TiO at $\sim$0.93 $\mu$m in the spectra of M dwarfs.

\end{abstract}

\keywords{infrared: stars --- stars:  late-type --- stars: low-mass, brown dwarfs ---  subdwarfs:  individual (LHS 3409, LHS 1135, LSR 2036$+$5059) --- stars:  individual (LSR 1610$-$0040)}

\section{Introduction}

In recent years a large population of objects with masses at and below the hydrogen burning minimum mass of $\sim$0.075 M$_{\odot}$ \citep[HBMM;][]{2001RvMP...73..719B} has been discovered \citep[e.g.,][]{1999ApJ...519..802K,2001A&A...380..590P,2004AJ....127.3553K}.  The majority of these objects were identified in wide-field surveys such as the Two Micron All Sky Survey \citep[2MASS;][]{1997ilsn.proc...25S}, the Deep Near Infrared Southern Sky Survey \citep[DENIS;][]{1997Msngr..87...27E}, and the Sloan Digital Sky Survey \citep[SDSS;][]{2000AJ....120.1579Y}.  These ``ultracool'' dwarfs  represent an extension of the near-solar-metallicity M dwarfs (which dominate the stellar population in the solar neighborhood in both number and mass) to lower masses and thus cooler effective temperatures.  Indeed the creation of two new spectral classes, `L' and `T', was required in order to properly classify these objects.

Despite the enormous increase in the number of catalogued ultracool dwarfs, only a limited number of ultracool \textit{subdwarfs} are known.  Late-type (FGKM) subdwarfs are metal-poor stars that lie to the left of the main sequence in the Hertzsprung-Russell (H-R) diagram \citep{1959MNRAS.119..278S}.  The position of these subdwarfs in the H-R diagram is a consequence of their reduced atmospheric opacity, which results in effective temperatures that are higher than those of their equivalent-mass main sequence counterparts.  Since these stars are typically members of the halo population, they have high space motions relative to the Sun and therefore are usually identified as high proper-motion objects \citep[e.g., ][]{1979lccs.book.....L}.  Due to the intrinsic faintness of M subdwarfs and their paucity in the solar neighborhood, only a few late-type M subdwarfs are known and as a result, current spectral classification schemes are limited to spectral types earlier than sdM7 \citep{1997AJ....113..806G}.  Nevertheless, $\sim$10 subdwarfs with spectral types that appear to be later than sdM7, including two L subdwarfs,  have recently been discovered with 2MASS \citep{2003ApJ...592.1186B,2004ApJ...614L..73B} and in new optical proper motion surveys \citep{2004A&A...421..763P,2005AJ....129.1483L}.

The first L-type subdwarf, 2MASS 0532$+$8246 (hereafter 2MASS 0532), was discovered serendipitously by \citet{2003ApJ...592.1186B} in a search for T dwarfs in the 2MASS database.  Its spectrum exhibits \ion{Rb}{1} and \ion{Cs}{1} lines, pressure-broadened \ion{Na}{1} and \ion{K}{1} lines, and FeH and CrH bands in the red optical, and H$_2$O bands in the near infrared, all of which are consistent with an L-type dwarf.  However it also exhibits CaH absorption, the deepest 0.986 $\mu$m FeH bandhead of any late-type dwarf known, and collision-induced H$_2$ 1$-$0 quadrupole absorption (CIA H$_2$) centered at 2.4 $\mu$m  \citep{2002A&A...390..779B} which heavily suppresses the flux in the $H$ and $K$ bands \citep{1994ApJ...424..333S,2000ApJ...535..965L}.  Enhanced hydride bands and strong H$_2$ absorption are indications of a metal-depleted, high-pressure atmosphere \citep{1982PASAu...4..417B,1994ApJ...424..333S,1995ApJ...445..433A}.   This, along with its high proper motion ($\mu$=2$\farcs$60 yr$^{-1}$) and radial velocity ($v_{\mathrm{rad}}$=$-$195 km s$^{-1}$), led \citet{2003ApJ...592.1186B} to classify 2MASS 0532 as an L-type subdwarf (sdL).

A second candidate L-type subdwarf, LSR 1610$-$0040 (hereafter LSR 1610), was discovered soon after by \citet{2003ApJ...591L..49L} in a search of the Digitized Sky Survey for high proper motion stars \citep{2002AJ....124.1190L}.  Its red-optical spectrum is atypical, exhibiting absorption features found in the spectra of both late-type dwarfs and subdwarfs.  Even though the red-optical spectrum of LSR 1610 does not fit into any standard dwarf or subdwarf sequence, it clearly exhibits spectral signatures consistent with having a very cool and metal-poor atmosphere and therefore \citet{2003ApJ...591L..49L} classified this object as an early-type L subdwarf.

Given the contradictory nature of the features in the red-optical spectrum of LSR 1610, and the fact that the near-infrared wavelength range is a powerful diagnostic of the gross metallicity of late-type dwarfs (e.g., CIA H$_2$, weak or absent CO), we have obtained a moderate resolution, 0.8$-$4.1 $\mu$m spectrum of LSR 1610 to re-examine the question of its spectral classification.  For comparison purposes, we have also obtained 0.8$-$2.4 $\mu$m spectra of LHS 3409 (sdM4.5), LHS 1135 (sdM6.5), and LSR 2036$+$5059 \citep[sdM7.5; ][]{2003AJ....125.1598L}.  In addition, our previous work \citep{2005ApJ...623.1115C} provides an infrared spectral sequence of field M and L dwarfs obtained with the same instrument against which the spectrum of LSR 1610 can also be compared. We discuss the observations and data reduction in \S2 while in \S3, we describe the red-optical and near-infrared spectrum of LSR 1610.  Finally in \S4, we discuss several possible explanations for the appearance of the spectrum of LSR 1610.

\section{Observations and Data Reduction}

Our observations were conducted with SpeX \citep{2003PASP..115..362R}, the facility near-infrared medium-resolution cross-dispersed spectrograph at the 3.0 m NASA Infrared Telescope Facility on Mauna Kea.  A log of the observations, including the UT date of the observations, spectroscopic mode, resolving power, total on-source integration time, and associated telluric standard stars, is given in Table \ref{obslog}.  Each target was observed in the short-wavelength cross-dispersed mode (SXD; 0.8$-$2.4 $\mu$m) with a 0$\farcs$3-wide slit which yielded a resolving power of $R$ $\equiv$ $\lambda/\Delta\lambda \approx$ 2000 in six spectral orders.  LSR 1610 was also observed in the long-wavelength cross-dispersed mode (LXD1.9; 1.9$-$4.1 $\mu$m) with a 0$\farcs$8-wide slit ($R$$\approx$940).  The observations consisted of a series of exposures taken at two different positions along the 15$''$-long slit.  An A0 V star was observed before or after each target to correct for absorption due to the Earth's atmosphere and to flux calibrate the science object spectra.  The airmass difference between each object and its associated ``telluric standard'' was always less than 0.05.  Finally, a set of internal flat field and arc exposures was taken after each object/standard pair for flat fielding and wavelength calibration purposes.

The data were reduced using Spextool \citep{2004PASP..116..362C}, the IDL-based data reduction package for SpeX.  The package performs nonlinearity corrections, flat fielding, optimal spectral extraction, and wavelength calibration.  The spectra were corrected for telluric absorption and flux calibrated ($\pm$10\%) using the extracted A0 V telluric standard and the technique described by \citet{2003PASP..115..389V}.  The spectra from the individual orders were then spliced together and the SXD and LXD1.9 spectra for LSR 1610 were combined.  In order to confirm that the order merging process did not introduce any errors in the relative flux density levels of the spectra, we have computed synthetic $J-H$ and $H-K_s$ 2MASS colors for the objects using the technique described in \citet{2005ApJ...623.1115C}.  We find that the synthetic colors of the spectra agree with the published 2MASS colors\footnote{The 2MASS magnitudes were obtained from \url{http://cdsweb.u-strasbg.fr/viz-bin/VizieR?-source=II/246}.} within the errors except for LHS 3409 whose synthetic $J-H$ color is 1.4-$\sigma$ redder than the 2MASS color.  Finally, the published red-optical (0.6$-$1.0 $\mu$m) spectrum of LSR 1610 (S. L\'{e}pine 2004, private communication) was combined with the SpeX spectrum resulting in a final spectrum covering from 0.6 to 4.1 $\mu$m.  We find that the overall slope of the SpeX spectrum from 0.8 to 1.0 $\mu$m  is shallower than that of the optical spectrum over the same wavelength range.  This discrepancy may be due to the fact that the standard stars used to calibrate the raw optical spectrum of LSR 1610 lack spectrophotometry between 8900 and 9700 \AA\,  due to H$_2$O telluric absorption \citep{1990ApJ...358..344M}.

\begin{figure}
\includegraphics[width=5in]{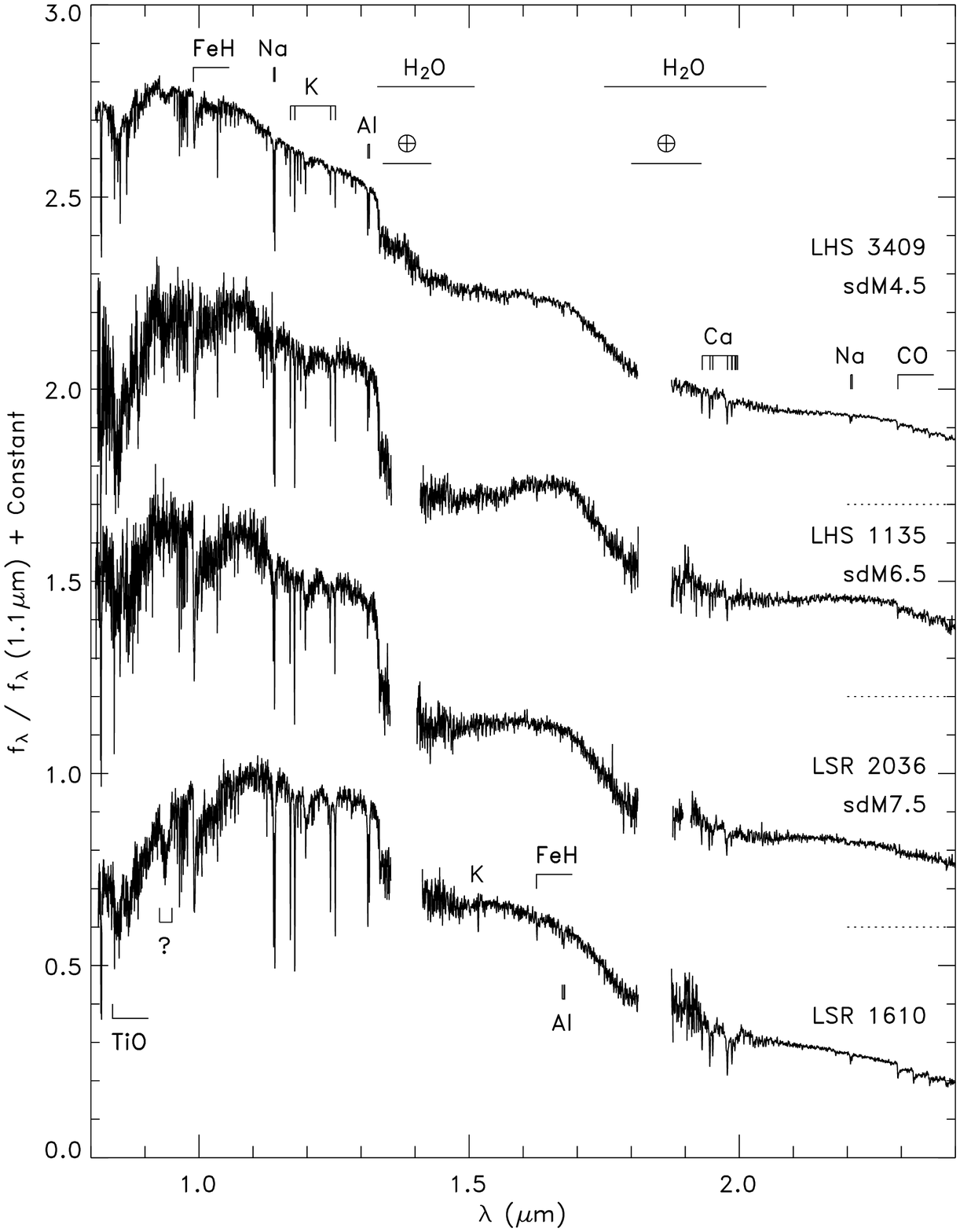}
\caption{\label{sdSeq}The 0.8$-$2.4 $\mu$m spectra of LHS 3409 (sdM4.5), LHS 1135 (sdM6.5), LSR 2036 (sdM7.5), and LSR 1610$-$0040.  The spectra have been normalized at 1.1 $\mu$m and offset by constants (\textit{dotted lines}).  The most prominent atomic and molecular absorption features are indicated.  The gap in the spectra at $\sim$1.85 $\mu$m is due to a gap in the wavelength coverage of the SXD mode of SpeX while the data were removed at 1.4 $\mu$m due to low a signal-to-noise ratio.}
\end{figure}

The 0.8$-$2.4 $\mu$m spectra of LSR 1610 and the three M subdwarfs are shown in Figure \ref{sdSeq}.  The signal-to-noise ratio (S/N) of the SXD spectra ranged from 50 up to 200 while the S/N of the LXD1.9 spectrum of LSR 1610 was $\sim$50.  We do not show the 2.4$-$4.1 $\mu$m spectrum of LSR 1610 because it is relatively featureless.  Absorption features commonly found in the spectra of late-type dwarfs and subdwarfs are indicated, including broad H$_2$O absorption bands centered at 1.4 and 1.8 $\mu$m, the Wing-Ford band of FeH  at 0.986 $\mu$m and strong \ion{Na}{1} and \ion{K}{1} lines in the $J$ band \citep[e.g., ][]{2000ApJ...535..965L,2004ApJ...614L..73B}.  Additional absorption features can also be discerned, including an \ion{Al}{1} doublet at $\sim$1.313 $\mu$m,  a \ion{Ca}{1} multiplet at $\sim$2$\mu$m, a \ion{Na}{1} doublet at $\sim$2.2 $\mu$m, overtone CO bands at $\sim$2.29 $\mu$m, a \ion{K}{1} line at 1.5167 $\mu$m,  an \ion{Al}{1} doublet at 1.6735 $\mu$m, and an FeH bandhead at  1.624 $\mu$m.

We have computed the bolometric flux of the LSR 1610 using the technique described in \citet{2005ApJ...623.1115C}.  Briefly, the 0.6$-$4.1 $\mu$m LSR 1610 spectrum was absolutely flux calibrated using 2MASS $J$-, $H$-, and $K_s$-band photometry \citep{2003ApJ...591L..49L}.  We then linearly interpolated from zero flux at zero wavelength to the flux density of the spectrum at 0.6  $\mu$m.  The gaps in the observed spectrum were removed by linearly interpolating from the gap edges and a Rayleigh-Jeans tail was assumed for $\lambda$ $>$ 4.1 $\mu$m.  Integrating over the spectrum yields $f_{\mathrm{bol}}$ = 2.72 $\pm$ 0.06 $\times$ 10$^{-14}$ W m$^{-2}$ ($m_{\mathrm{bol}}$=14.92 $\pm$ 0.03)\footnote{$m_{\mathrm{bol}}  = -2.5 \times \log(f_{\mathrm{bol}}) - 18.988$ assuming $L_{\odot} = 3.86\times10^{26}$ W and $M_{\mathrm{bol}\odot} = +4.74$.}.  We made no attempt to compute the bolometric fluxes of the M subdwarfs because the spectra cover only from 0.8 to 2.4 $\mu$m.

\section{The Spectrum of LSR 1610$-$0040}

\subsection{The Red-Optical Spectrum}

\citet{2003ApJ...591L..49L} classified LSR 1610 as an early-type L subdwarf based primarily on three lines of evidence drawn from its red-optical spectrum.  Firstly, the spectrum exhibits features present in L dwarf spectra including \ion{Rb}{1} lines, a CrH bandhead, and a  FeH bandhead.  Secondly, the spectrum contains features indicative of a metal-poor atmosphere including strong CaH absorption, weak TiO bandheads, and weak VO absorption.  Finally, the spectrum of LSR 1610 is significantly redder than the  late-type M subdwarf, LSR 1425$+$7102 \citep[sdM8; hereafter LSR 1425; ][]{2003ApJ...585L..69L} which indicates that its spectral type is later than sdM8.  Interestingly, \citet{2003ApJ...591L..49L} derive a spectral type of sdM6 using the CaH2, CaH3, and TiO5 spectral indices developed by \citet{1997AJ....113..806G} that measure the strengths of CaH and TiO bands near 7000 \AA.

\begin{figure}
\includegraphics[width=5in]{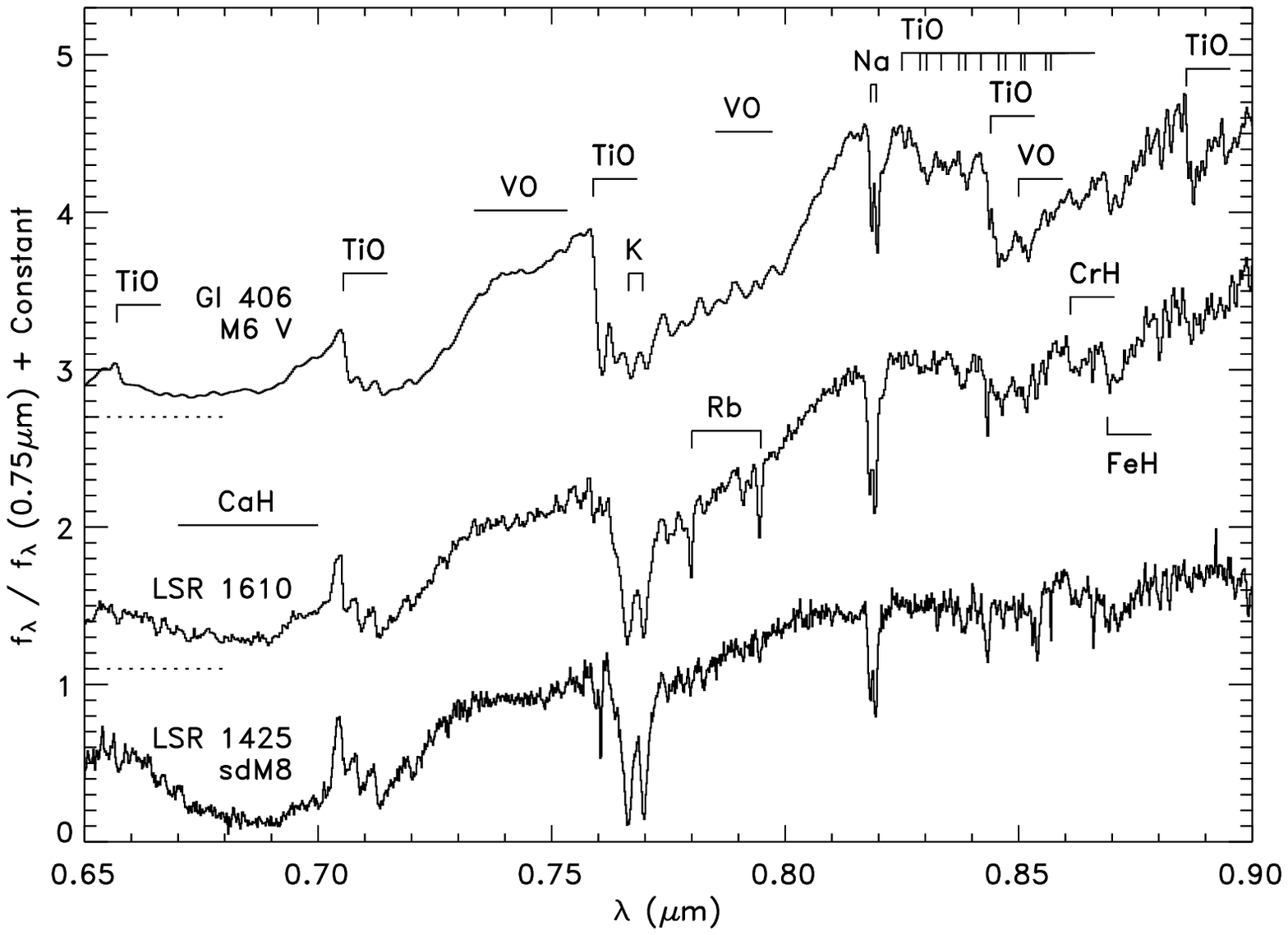}
\caption{\label{optLSR}The 0.65$-$0.9 $\mu$m spectra of Gl 406 \citep[M6 V; ][]{1991ApJS...77..417K,2005ApJ...623.1115C}, LSR 1610 \citep{2003ApJ...591L..49L} and LSR 1425$+$7102 \citep[sdM8; ][]{2003ApJ...585L..69L}.  The spectra have been normalized at 0.75 $\mu$m and offset by constants (\textit{dotted lines}).  The most prominent atomic and molecular absorption features are indicated.}
\end{figure}

Figure \ref{optLSR} shows the 0.6$-$0.9 $\mu$m spectrum of LSR 1610 \citep{2003ApJ...591L..49L}.  Also shown for comparison purposes are the spectra of a late-type M dwarf, Gl 406 \citep[M6 V; ][]{1991ApJS...77..417K,2005ApJ...623.1115C}, and a late-type M subdwarf LSR 1425 \citep{2003ApJ...585L..69L}. The spectrum of LSR 1425 is clearly a better overall match to that of LSR 1610 than the spectrum of Gl 406.  In particular, the TiO bandheads at 6569, 7589, 8432, and 8859 \AA, which are quite prominent in the spectrum of Gl 406, are weak or absent in the spectra of LSR 1425 and LSR 1610.  Due to the loss of the overlying TiO opacity, the resonance \ion{K}{1} doublet (7665, 7699 \AA), which is only weakly detected in the spectrum of Gl 406, is also prominent in the spectra of LSR 1425 and LSR 1610.  Finally the spectra of both LSR 1425 and LSR 1610 exhibit strong CaH absorption from 6750 to 7050 \AA.  All of these features suggest LSR 1610 is a subdwarf.  Despite the good agreement between the red-optical spectra of LSR 1425 and LSR 1610, the spectrum of LSR 1610 also exhibits features that are usually associated with L dwarfs including a resonant \ion{Rb}{1} doublet (7800, 7948 \AA) and CrH (8611 \AA) and FeH (8692 \AA) bandheads \citep{1999ApJ...519..802K}.  Nevertheless, \citet{2004A&A...428L..25S} have shown that the spectrum of SSSPM J1444$-$2019, a candidate L subdwarf, also exhibits these features which suggests their presence may not be uncommon in the spectra of late-type subdwarfs.  In summary, the red-optical spectrum of LSR 1610 is more similar to the spectrum of the M subdwarf LSR 1425 than to the M dwarf Gl 406.

\subsection{The Near-Infrared Spectrum}

\begin{figure}
\includegraphics[width=5in]{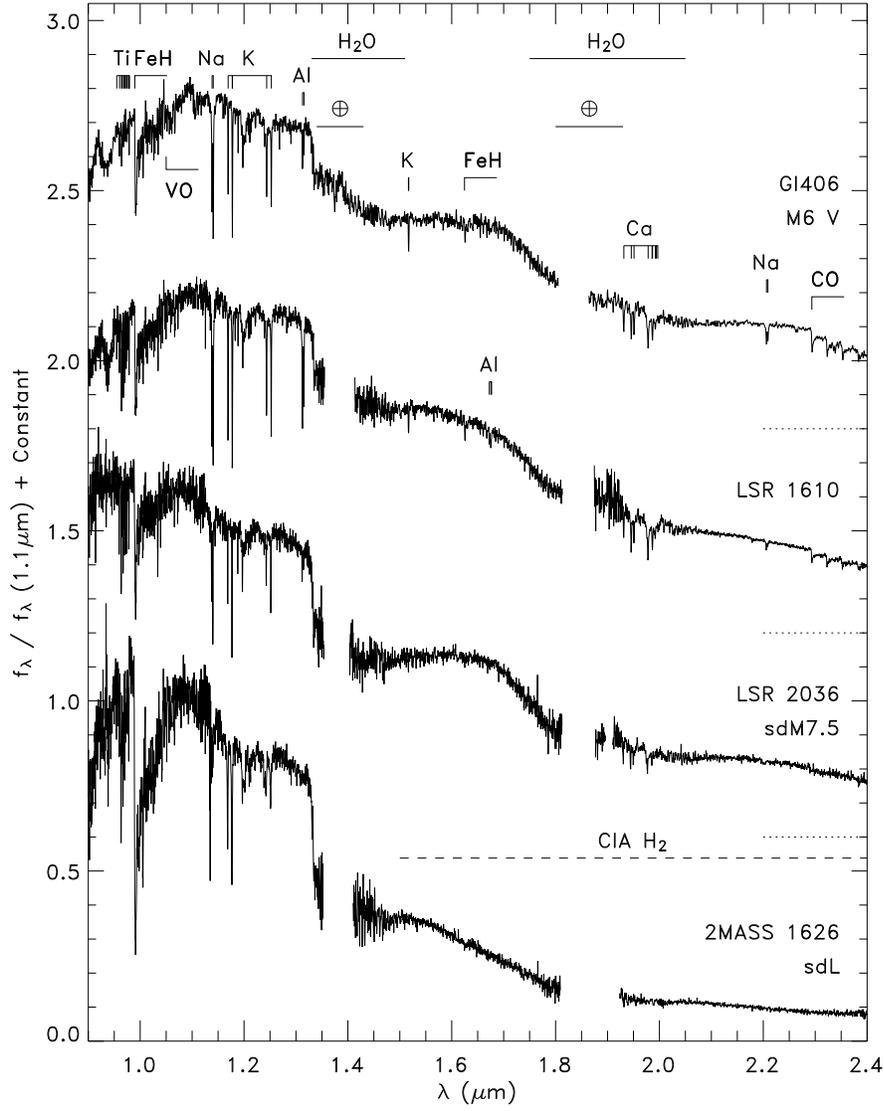}
\caption{\label{irLSR}The 0.9$-$2.4 $\mu$m spectra of Gl 406 \citep[M6 V; ][]{2005ApJ...623.1115C}, LSR 1610, LSR 2036 (sdM7.5), and 2MASS 1626$+$3925 \citep{2004ApJ...614L..73B}.  The spectra have been normalized at 1.1 $\mu$m and offset by constants (\textit{dotted lines}).  The most prominent atomic and molecular absorption features are indicated.}
\end{figure}

Based on the properties of the red-optical spectrum, we would expect the near-infrared spectrum of LSR 1610 to exhibit features similar to those of other late-type M and L subdwarfs.  Figure \ref{irLSR} shows the 0.9$-$2.4 $\mu$m spectra of LSR 1610, along with the spectrum of Gl 406 \citep[M6 V;][]{2005ApJ...623.1115C}, LSR 2036 (sdM7.5), and the SpeX SXD spectrum of 2MASS 1626$+$3925 \citep[hereafter 2MASS 1626, ][]{2004ApJ...614L..73B}, an early-type L subdwarf.  Figures \ref{sdSeq} and \ref{irLSR} clearly show that LSR 1610 does not fit neatly into the M/L subdwarf sequence.  Instead, we find that it exhibits a mix of characteristics found in the spectra of both M dwarfs and M subdwarfs.

As shown in Figure 3, the near-infrared spectrum of the M dwarf Gl 406 is actually a better match to that of LSR 1610 than either the spectrum of LSR 2036 or 2MASS 1626.  In fact, of all the M dwarfs in the \citet{2005ApJ...623.1115C} sample, Gl 406 provides the best overall match to the 0.6$-$4.1 $\mu$m spectrum of LSR 1610.  In particular, the spectrum of LSR 1610 and Gl 406 both peak at $\sim$1.1 $\mu$m while the spectra of the subdwarfs do so shortward of $\sim$1 $\mu$m.     In addition, the spectra of Gl 406 and LSR 1610 both exhibit overtone CO bandheads and a \ion{Na}{1} doublet in the $K$ band, and a \ion{K}{1} line and FeH bandhead in the $H$ band.  All of these features are either absent or weak in the spectra of late-type M and L subdwarfs which, taken at face value, suggests that LSR 1610 is an M dwarf.  Finally, \citet{2004ApJ...614L..73B} found that the \ion{K}{1} doublet lines in the $J$-band spectra of 2MASS 1626 and 2MASS 0532 are $\sim$20\% broader than in the spectra of field L dwarfs of similar spectra types (Kelu-1AB (L2) and DENIS 0205$-$1159AB (L7), respectively).  This effect is presumably due to pressure broadening since the reduced metal abundances of subdwarfs allows observations of deeper, and thus higher pressure, layers of the atmosphere.  We measured the FWHM of the short-wavelength \ion{K}{1} doublet in the spectrum of LSR 1610 to be 8.47 and 9.83 \AA\,, 30$-$35\% smaller than the values measured for the same doublet in the spectrum of Kelu-1AB \citep{2005ApJ...623.1115C}.  This provides further evidence that LSR 1610 is not an L subdwarf.  These FWHM values are, however, consistent with those measured for Gl 406 (M6 V) and are 25$-$30\% larger than those measured for LSR 2036 (sdM7.5).

\begin{figure}
\includegraphics[width=5in, angle=90]{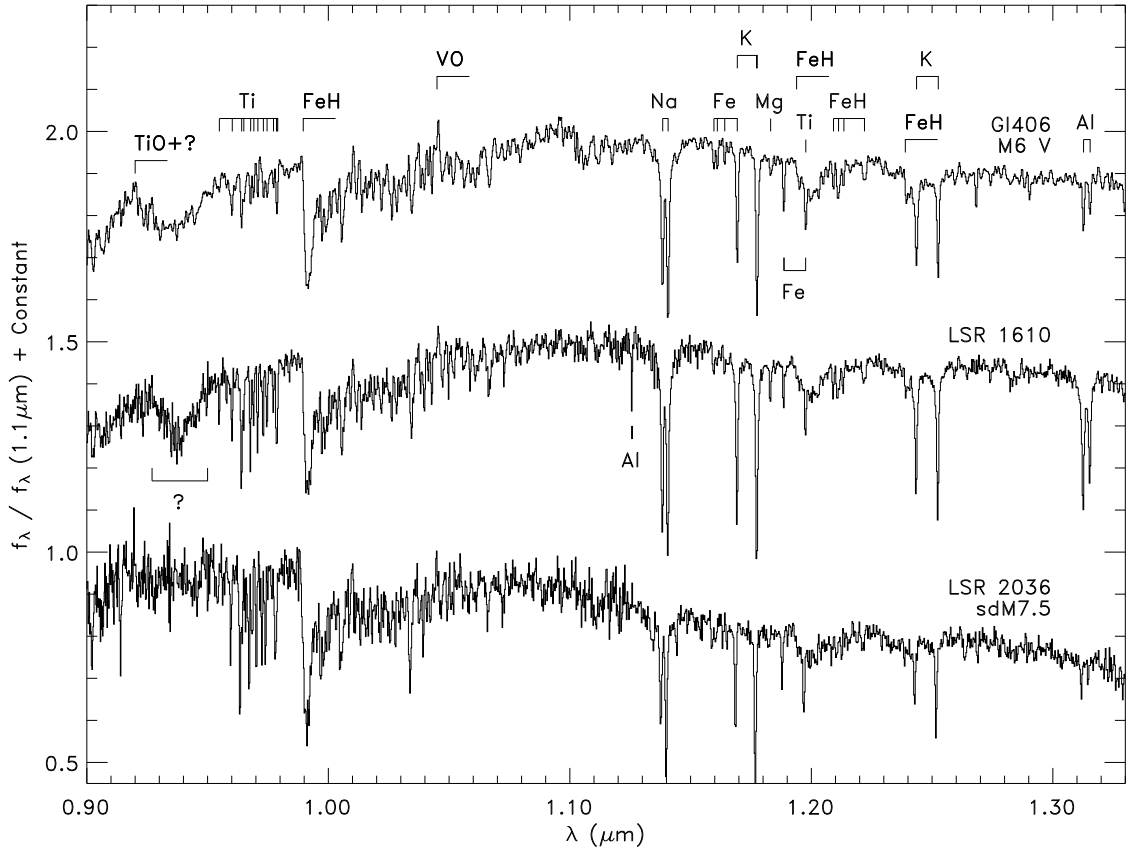}
\caption{\label{Jband}The 0.9$-$1.34 $\mu$m spectra of Gl 406 \citep[M6 V; ][]{2005ApJ...623.1115C}, LSR 1610, and LSR 2036 (sdM7.5).  The spectra have been normalized at 1.1 $\mu$m and offset by constants.  Prominent absorption features due to FeH, Na, K, Al, and Fe are indicated. Note the strong Al doublet at $\sim$1.315 $\mu$m and the \ion{Al}{1} line at 1.12 $\mu$m in the spectrum of LSR 1610.}
\end{figure}

Despite the fact that the near-infrared spectrum of LSR 1610 looks superficially like an M dwarf, it also exhibits features seen in the spectra of M subdwarfs.  As can be seen in Figure \ref{irLSR}, the spectrum of LSR 1610 appears flattened relative to that of Gl 406 in the $H$ and $K$ bands, though not as much as 2MASS 1626.  The relative flux density levels of the two spectra agree at both longer and shorter wavelengths which indicates an additional opacity source at the $H$ and $K$ bands may be present in the atmosphere of LSR 1610.  The most obvious candidate is collision-induced H$_2$ absorption centered at 2.4 $\mu$m  \citep{2002A&A...390..779B} which suppresses the flux in the $H$- and $K$-band spectra of low-metallicity dwarfs \citep{1994ApJ...424..333S,2000ApJ...535..965L}.  However since CIA H$_2$ lacks any distinct spectral features, this identification remains uncertain.  As shown in Figure \ref{Jband}, the spectra of LSR 1610 and LSR 2036 lack the 0$-$0 band of the $A$ $^4\Pi- X$ $^4\Sigma^-$ transition of VO centered at $\sim$1.06 $\mu$m which is prominent in the spectra of late-type M dwarfs \citep{2005ApJ...623.1115C}.  The absence of this band is consistent with the weak VO bands seen in the red-optical spectrum of LSR 1610.  Finally, the lines composing the $a$ $^5$F $-$ $z$ $^5$F$^\circ$ \ion{Ti}{1} multiplet centered at 0.97 $\mu$m are stronger in the spectrum of LSR 1610 than in typical M dwarfs (e.g., Gl 406);  the line strengths are more consistent with those in the spectra of M subdwarfs (e.g., LSR 2036).

\begin{figure}
\includegraphics[width=5in]{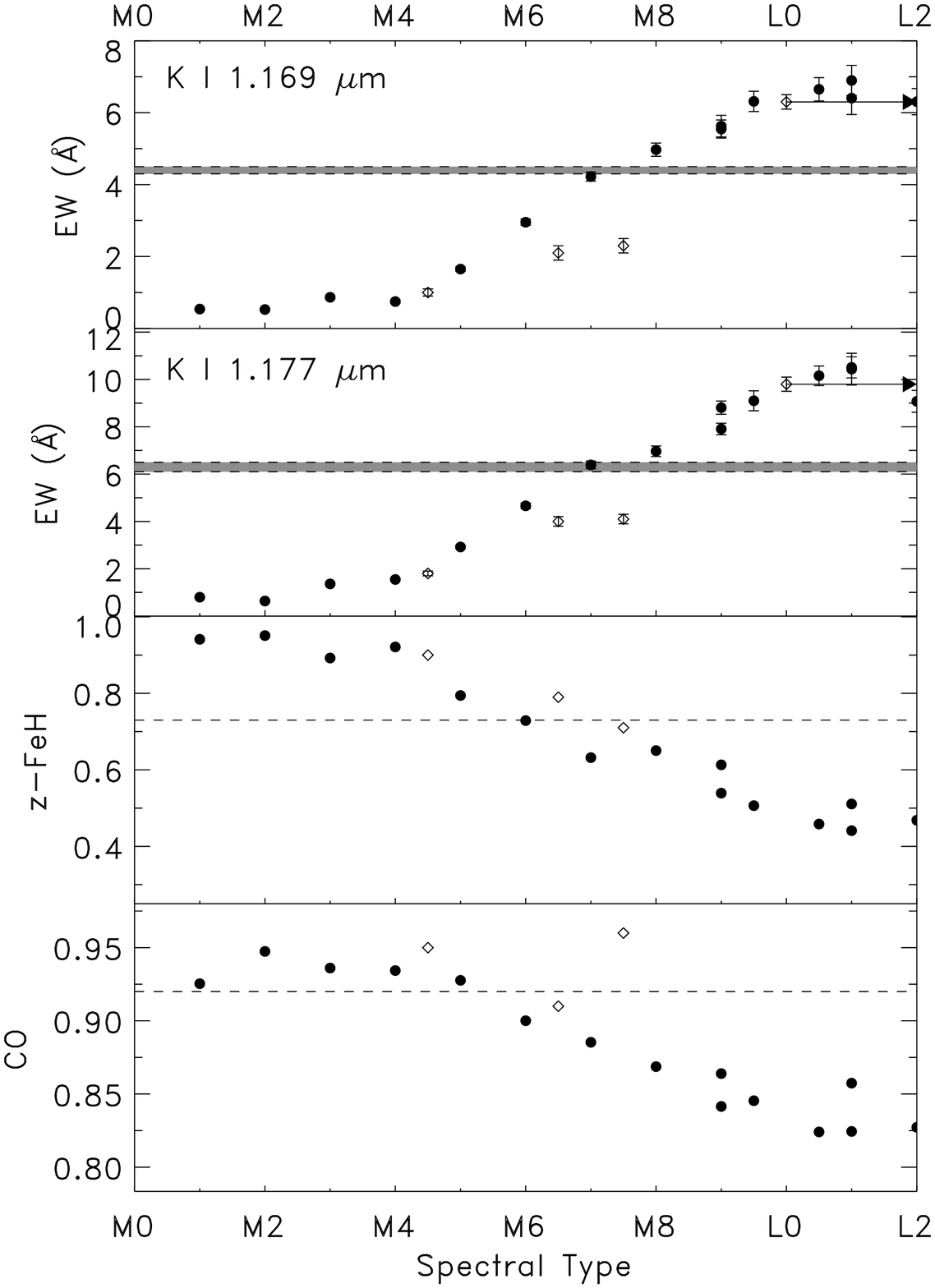}
\caption{\label{pEWs} The EWs of the 1.169 and 1.177 $\mu$m \ion{K}{1} lines and the CO and $z$-FeH flux ratios of \citet{2003ApJ...596..561M} as a function of spectral type.  The M dwarfs from the \citet{2005ApJ...623.1115C} sample are shown as filled circles while the M subdwarfs in the current sample are shown as diamonds.  The EW of 2MASS 1626$+$2925 is shown as a lower limit since its spectral subclass is unknown.  The values for LSR 1610 are indicated by dashed lines; the greyed region denotes the $\pm$1$\sigma$ EW range.}
\end{figure}

Since the spectrum of LSR 1610 exhibits features found in both M dwarf and M subdwarf spectra, and does not match the spectral energy distribution of known L dwarfs and L subdwarfs, it is clear that based on its gross near-infrared properties, LSR 1610 has a spectral type of M rather than L.  To further constrain its spectral type, we have computed the equivalent widths (EWs) of the short wavelength \ion{K}{1} doublet (1.169/1.177 $\mu$m) following the technique described in \citet{2005ApJ...623.1115C} for LSR 1610 and the M subdwarfs in our sample.  We have also computed the \citet{2003ApJ...596..561M} CO and $z$-FeH flux ratios that measure the strengths of the 2.29 $\mu$m CO bandhead and the 0.986 $\mu$m FeH bandhead, respectively.  The results are given in Table \ref{tEWs}.  Figure \ref{pEWs} shows the \ion{K}{1} EWs and CO and $z$-FeH flux ratios of LSR 1610 and the M subdwarfs as a function of spectral type.  Also shown are the values for the M dwarfs in the \citeauthor{2005ApJ...623.1115C} sample.  If we assume LSR 1610 is a dwarf, the EWs of the \ion{K}{1} lines and the spectral ratios indicate LSR 1610 has a spectral type of M5$-$M7 V, in agreement with the spectral type of $\sim$M6 V estimated from the overall spectral energy distribution of LSR 1610.  We note, however, that the EWs of the \ion{K}{1} lines and the strength of the FeH bandhead are also consistent with LSR 1610 being an M subdwarf with a spectral type later than sdM7.5.  Furthermore, the strength of the CO bandhead is also consistent with LSR 1610 being an M subdwarf with a spectral type of $\sim$sdM6.

Finally there are also two features in the spectrum that are inconsistent with our tentative classification of LSR 1610 as an M dwarf/subdwarf.  The first is the strength of the \ion{Al}{1} lines.  As can be seen in Figure \ref{Jband}, the doublet at 1.313 $\mu$m is much stronger in the spectrum of LSR 1610 than in either the spectrum of Gl 406 or LSR 2036.  We have computed the EW of this doublet for LSR 1610 and the M subdwarfs in our sample and the results are given in Table \ref{tEWs}.  The EW of the doublet in the spectrum of LSR 1610 is a factor of $\sim$3 greater than that measured for the three M subdwarfs and typical M dwarfs \citep{2005ApJ...623.1115C}.  We also note that the \ion{Al}{1} doublet is absent in the spectra of L dwarfs \citep{2003ApJ...596..561M,2005ApJ...623.1115C} and L subdwarfs \citep{2004ApJ...614L..73B}.  Additional \ion{Al}{1} lines at 1.1259 $\mu$m (see Figure \ref{Jband}) and 1.674 $\mu$m are also present in the spectrum of LSR 1610.  These lines are detected only in the spectra of M dwarfs with spectral types earlier than M5 V \citep{2005ApJ...623.1115C} and in only one of the three M subdwarfs in our sample (LHS 3409, sdM4.5).

The second feature is an unidentified, triangular-shaped, absorption band centered at 0.9375 $\mu$m and extending from 0.927 to 0.950 $\mu$m.  As can be seen in Figure \ref{Jband}, the spectrum of LSR 2036 lacks an absorption band at this wavelength while the spectrum of Gl 406 exhibits a broader, more bowl-shaped absorption feature centered at 0.933 $\mu$m and extending from 0.920 to 0.952 $\mu$m.  We have compared  the spectrum of LSR 1610 to the absorption cross-section spectra of a number of species with bands near 0.9375 $\mu$m including TiO, TiH, H$_2$O, and VO in attempt to determine the carrier of the unidentified band.  The cross-section spectra were kindly provided by R. Freedman (2005, private communication) and were computed at $T$=2000 K and $P$=1 bar.  None of these species has a band that matches the position, depth, or shape of the band in the spectrum of LSR 1610.  In the appendix, we show that the absorption band at 0.933 $\mu$m in the spectra of M dwarfs is carried, at least in part, by TiO.  In particular, the absorption from 0.920 to 0.927 $\mu$m can be entirely accounted for by TiO.  The lack of absorption at these wavelengths in the spectrum of LSR 1610 is consistent with the weak TiO bands seen in the red-optical spectrum.

\section{Discussion and Conclusions}

Table \ref{Features} summarizes the spectral features identified in the spectrum of LSR 1610 along with the best estimate of the spectral type of LSR 1610 obtained from each feature.  The red-optical spectrum clearly suggests an M subdwarf spectral type while the near-infrared spectrum suggests and M dwarf spectral type.  It therefore seems that the spectrum of LSR 1610 does not fit into either the tentative M/L subdwarf sequence or the well-established M/L dwarf sequence.

What then to make of LSR 1610?  There are a number of possibilities that might explain the contradictory spectral features.

\textit{M Dwarf with Low Oxygen Abundance}-- If we posit that LSR 1610 is an M dwarf with an overall metallicity near solar, then the features most at odds with this classification are the weak TiO and VO bands in the red-optical.  The weakness of these bands and the strengths of the \ion{Ti}{1} lines at $\sim$0.97 $\mu$m could simultaneously be explained by assuming a low oxygen abundance relative to solar; the abundance of TiO and VO would decrease due to the loss of O and the relative abundance of the gas-phase Ti would consequently increase \citep{2002ApJ...577..974L}.  However the strengths of the H$_2$O and CO bands in the near-infrared are not atypical for M dwarfs/subdwarfs and thus eliminate this possibility.

\textit{M Dwarf with Enhanced Grain Formation}--  The TiO and VO bands weaken across the M$\rightarrow$L dwarf transition \citep{1997ApJ...480L..39J,1999ApJ...519..802K} due, in part, to the formation of titanium- and vanadium-bearing condensates \citep{1999ApJ...512..843B,2002ApJ...577..974L}.  The M$\rightarrow$L transition occurs at $T_{\mathrm{eff}}$ $\simeq$ 2300 K \citep{2004AJ....127.3516G} while LSR 1610 has $T_{\mathrm{eff}}$ $\simeq$ 2700 K if we assume it has a spectral type of M6 V (see \S3.2).  This indicates that any condensation would have to be significantly enhanced.  The presence and strength of the \ion{Ti}{1}, \ion{Ca}{1}, \ion{Mg}{1}, and \ion{Al}{1} lines in the spectrum of LSR 1610 belie this hypothesis since we would expect that these refractory elements would also be removed from the atmosphere by condensation \citep{1999ApJ...512..843B,2002ApJ...577..974L}.

\textit{Peculiar M Subdwarf}-- LSR 1610 could be a peculiar M subdwarf.   In this case, then the presence of the \ion{Na}{1} doublet and CO overtone bands in the $K$ band, and the \ion{K}{1} line, FeH bandhead, and \ion{Al}{1} doublet in the $H$ band must be explained.  One possibility is that these species are overabundant.  However the strengths of \ion{Na}{1} and \ion{K}{1} lines in the $J$ band and the FeH bandhead at 0.986 $\mu$m are not anomalous.  An alternative hypothesis is that the collision-induced H$_2$ absorption in the $H$ and $K$ bands is unusually weak allowing these features to emerge.  Since the strength of the CIA H$_2$ is sensitive to the atmospheric pressure, and the atmospheric pressure is, all else being equal, proportional to the surface gravity of the object, this may indicate LSR 1610 has a low surface gravity.  However the $H$ and $K$ band spectrum of LSR 1610 appears flatter than the spectrum of LSR 2036 (sdM7.5)  which argues against this hypothesis.

\textit{M/L Dwarf Binary}--  Since LSR 1610 exhibits features seen in the spectra of M and L dwarfs, could it be an unresolved M/L binary?  We find this hypothesis unlikely for three reasons.  Firstly, L dwarfs are fainter than M dwarfs at both optical and near-infrared wavelengths which would make the detection of the \ion{Rb}{1} lines and FeH and CrH bandheads in a composite M/L spectrum unlikely.  Secondly, the optical spectrum of LSR 1610 lacks the 8521 \AA\, \ion{Cs}{1} line, which is a distinctive feature in the spectra of L dwarfs \citep{1999ApJ...519..802K}.  Finally, the overall red-optical spectrum of LSR 1610 is completely inconsistent with that of M and L dwarfs \citep{1999ApJ...519..802K}.

\textit{Mildly Metal-Poor M Dwarf}-- The overall spectral energy distribution of LSR 1610 is clearly best matched by that of Gl 406, an M6 dwarf.  If we therefore assume that LSR 1610 is a mid-type M dwarf, we can explain many of the seemingly disparate spectral features if we further posit that is it mildly metal poor.  A metallicity intermediate between M dwarfs and M subdwarfs would explain the subdwarf-like spectral features (weak TiO and VO bands, mild CIA H$_2$, and prominent \ion{Ti}{1} lines) and the dwarf-like spectral features (CO bands, \ion{Na}{1} doublet, 1.617 $\mu$m FeH bandhead, and 1.516 $\mu$m \ion{K}{1} line).  Given the information currently available for LSR 1610, we find this to be the most plausible explanation for the features seen in its spectrum.

Nevertheless, none of the above scenarios can explain the strengths of the \ion{Al}{1} lines in the spectrum of LSR 1610 and in particular, the anomalously large EW of the \ion{Al}{1} doublet at 1.313 $\mu$m.  Since the strength of the H$_2$O absorption that provides the continuum at this wavelength appears normal, we conclude \ion{Al}{1} must be overabundant relative to solar.  In addition, there remains the unidentified molecular absorption band centered at 0.9375 $\mu$m that is completely inconsistent with the spectra of known M dwarfs and M subdwarfs.  Since the parallax of LSR 1610 has yet to be measured, it is difficult to further constrain its nature.  It is, however, currently being observed as part of the United States Naval Observatory (USNO) faint star parallax program (H. Harris 2005, private communication). The forthcoming parallax measurement can then be combined with our $f_{\mathrm{bol}}$ measurement (see \S2) to compute its bolometric luminosity which may shed further light on the nature of LSR 1610.

\acknowledgments

The authors wish to thank S. L{\' e}pine for providing the optical spectrum of LSR 1610$-$0040, A. Burgasser for providing the spectrum of 2MASS 1626$+$3925, R. Freedman for providing the molecular cross-section data, A. Burgasser and J. Liebert for stimulating discussions, and the anonymous referee whose suggestions improved the paper.  This publication makes use of data from the Two Micron All Sky Survey, which is a joint project of the University of Massachusetts and the Infrared Processing and Analysis Center, and funded by the National Aeronautics and Space Administration and the National Science Foundation.  This research has made use of the SIMBAD database, operated at CDS, Strasbourg, France and NASA's Astrophysics Data System Bibliographic Services.  M.C.C. acknowledges financial support from the NASA Infrared Telescope Facility and NASA through the Spitzer Space Telescope Fellowship Program.

\appendix

\section{The Tentative Identification of a New TiO Band in M Dwarf Spectra}

In the course of attempting to identify the absorption band at 0.9375 $\mu$m in the spectrum of LSR 1610, we found that the identification of H$_2$O as the primary carrier of an absorption band centered at $\sim$0.933 $\mu$m in the spectra of M dwarfs by \citet{2005ApJ...623.1115C} was in error.  That identification was based on the apparent coincidence between the position of the absorption band in the spectra of M dwarfs and the telluric H$_2$O band.  In this appendix, we show that this band is carried, at least in part, by TiO.

\begin{figure}
\includegraphics[width=5in]{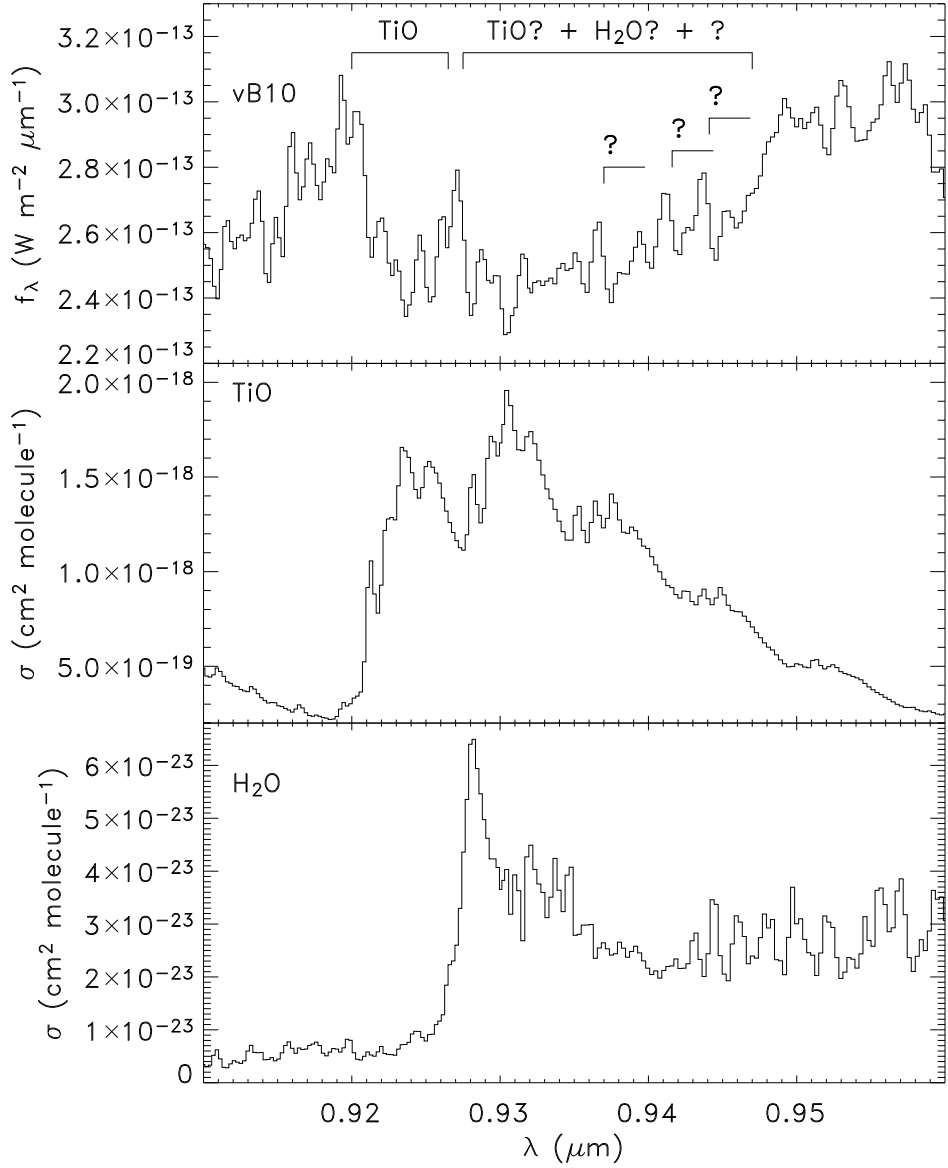}
\caption{\label{TiO}\textit{Bottom}: Cross-section spectrum of H$_2$O at $T$=2000 K and $P$=1 bar.  \textit{Middle}:  Cross-section spectrum of TiO at $T$=2000 K and $P$=1 bar.  \textit{Top}:  Spectrum of vB 10 \citep[M8 V;][]{2005ApJ...623.1115C}.  The broad absorption band in the spectrum of vB 10 is carried, at least in part, by TiO.}
\end{figure}

Figure \ref{TiO} shows the $R$ $\approx$ 2000 spectrum of a typical late-type M dwarf vB 10 \citep[M8 V;][]{2005ApJ...623.1115C} in the upper panel and the cross-section data of TiO and H$_2$O in the middle and lower panels, respectively.  The spectrum of vB 10 has a S/N $>$ 100.   The cross-section data, which were computed at $T$=2000 K and $P$=1 bar (R. Freedman 2005, private communication), have been smoothed to $R$=2000 and resampled onto the wavelength grid of vB 10.  The absorption band, which covers from 0.92 to $\sim$0.948 $\mu$m, is clearly inconsistent with the H$_2$O cross-section spectrum.  However the TiO data matches the width of the band relatively well.

The absorption band actually appears to have two distinct parts, ranging from 0.92 to 0.927 $\mu$m and 0.927 to 0.948 $\mu$m.  The 0.92 to 0.948 $\mu$m section of the band can be accounted for by TiO but the rest of the band is probably carried by a combination of TiO, H$_2$O and some as-yet unidentified species.  Indeed there appears to be additional bandheads at 0.9370, 0.9416, and 0.9441 $\mu$m that are absent in the TiO or H$_2$O data.  Higher resolution spectra will be required to both confirm the identification of TiO as a partial carrier of this band and identify any additional carriers.

\bibliographystyle{apj}
\bibliography{ref,tmp}

\clearpage

\clearpage

\clearpage

\clearpage

\clearpage

\clearpage

\input{tab1.tex}

\clearpage

\input{tab2.tex}

\clearpage

\input{tab3.tex}

\end{document}

%% file: tab1.tex
\begin{deluxetable}{lllllcl}
\tablecolumns{10}
\tabletypesize{\scriptsize} 
\tablewidth{0pc}
\tablecaption{\label{obslog}Log of SpeX Observations}
\tablehead{
\colhead{Object} & 
\colhead{Spectral\tablenotemark{a}} & 
\colhead{UT Date} & 
\colhead{Spectroscopic} &
\colhead{$R$} &
\colhead{Exp. Time} & 
\colhead{A0 V} \\

\colhead{} & 
\colhead{Type} &
\colhead{} & 
\colhead{Mode} & 
\colhead{} & 
\colhead{(sec)} & 
\colhead{Standard}}

\startdata

LHS 3409        & sdM4.5  & 2003$-$07$-$06 & SXD    & 2000 & 1440 & HD 178207  \\
LHS 1135        & sdM6.5  & 2003$-$10$-$15 & SXD    & 2000 & 3000 & HD 7215  \\
LSR 2036$+$5059 & sdM7.5  & 2003$-$10$-$06 & SXD    & 2000 & 2400 & HD 205314 \\
LSR 1610$-$0040 & sdL?    & 2003$-$07$-$06 & SXD    & 2000 & 1680 & HD 140775  \\
                &         & 2003$-$07$-$06 & LXD1.9 & \phn940  & 2400 & HD 140775 \\

\enddata

\tablenotetext{a}{The spectral types are based on optical spectroscopy and are from \citet{1997PASP..109..849G} and \citet{2003ApJ...591L..49L,2003AJ....125.1598L}.}

\end{deluxetable}

%% file: tab2.tex
\begin{deluxetable}{llccccc}
\tabletypesize{\scriptsize}
\tablecolumns{5}
\tablewidth{0pc}
\tablecaption{\label{tEWs}Equivalent Widths and Flux Ratios of LSR 1610$-$0040}
\tablehead{

\colhead{Object} & 
\colhead{Spectral Type} & 
\multicolumn{3}{c}{{EW (\AA)}} \\

\cline{3-5}

\colhead{} & 
\colhead{} & 
\colhead{\ion{K}{1} (1.169 $\mu$m)} &
\colhead{\ion{K}{1} (1.177 $\mu$m)} &
\colhead{\ion{Al}{1} (1.132 $\mu$m)} &
\colhead{$z$-FeH} &
\colhead{CO} \\

\colhead{(1)} &
\colhead{(2)} &
\colhead{(3)} &
\colhead{(4)} &
\colhead{(5)} &
\colhead{(7)} &
\colhead{(8)}}

\startdata

LHS 3409          & sdM4.5   &  1.0 $\pm$ 0.1 &  1.8 $\pm$ 0.1 &  2.5 $\pm$ 0.1   & 0.90 & 0.95 \\
LHS 1135          & sdM6.5   &  2.1 $\pm$ 0.2 &  4.0 $\pm$ 0.2 &  2.7 $\pm$ 0.2   & 0.79 & 0.91 \\
LSR 2036$+$5059   & sdM7.5   &  2.3 $\pm$ 0.2 &  4.1 $\pm$ 0.2 &  1.4 $\pm$ 0.2   & 0.71 & 0.96 \\
LSR 1610$-$0040   & sdL?     &  4.4 $\pm$ 0.1 &  6.3 $\pm$ 0.2 &  8.5 $\pm$ 0.2   & 0.73 & 0.92 \\


\enddata

\end{deluxetable}

%% file: tab3.tex
\begin{deluxetable}{llll}
\tablecolumns{4}
\tabletypesize{\scriptsize} 
\tablewidth{0pc}
\tablecaption{\label{Features}Summary of Features in the Spectrum of LSR 1610$-$0040}
\tablehead{
\colhead{} & 
\colhead{} & 
\colhead{} &
\colhead{Corresponding} \\

\colhead{Feature} & 
\colhead{Wavelength ($\mu$m)} & 
\colhead{Notes} &
\colhead{Spectral Type}}

\startdata

CaH band      & 0.6750$-$0.7050   & Strongest in sdM                                        & sdM6      \\
TiO bandhead  & 0.7053            & Strongest in late dM and sdM                            & M         \\
VO band       & 0.7534$-$0.7734   & Strongest in late dM and early dL                       & sdM       \\
TiO bandhead  & 0.7589            & Strongest in late dM and early dL                       & sdM       \\
\ion{K}{1}    & 0.7665,0.7699     & Seen in dM, dL and sdL                                  & sdM       \\
VO band       & 0.7851$-$0.7973   & Strongest in late dM and early dL                       & sdM         \\
\ion{Rb}{1}   & 0.7800,0.7948     & Seen in dL                                              & L         \\
\ion{Na}{1}   & 0.8183, 0.8195    & Seen in dM, early dL, and sdM                           & $\cdots$       \\
TiO bandhead  & 0.8206            & Strongest in late dM and early dL                       & sdM       \\
TiO bandhead  & 0.8432            & Strongest in dM and early dL                            & sdM       \\
CrH bandhead  & 0.8611            & Seen in dL and sdL                                      & L         \\
FeH bandhead  & 0.8692            & Seen in dL and sdL                                      & L         \\
\ion{Ti}{1}   & 0.96$-$0.98       & Seen in dM.  Strongest in sdM                           & sdM       \\
FeH bandhead  & 0.9896            & Seen in dM, dL, sdM, and sdL                            & $\sim$M6 V, $\sim$sdM7  \\
\ion{Al}{1}   & 1.126             & Seen only in early to mid dM and sdM                    & $<$M5   \\
\ion{Na}{1}   & 1.138,1.140       & Seen in dM, dL and sdM                                  & M, L   \\
\ion{K}{1}    & 1.169,1.178       & Seen in dM and dL                                       & $\sim$M7 V, $\sim$sdM8  \\
\ion{Al}{1}   & 1.313             & Seen in dM and sdM                                      & $\cdots$    \\
\ion{K}{1}    & 1.516             & Seen in dM and dL and weakly in sdM                     & dM  \\
FeH bandhead  & 1.62457           & Seen in dM and dL and weakly in sdM                     & dM  \\              
\ion{Al}{1}   & 1.674             & Seen in early dM and sdM                                & dM  \\              
\ion{Ca}{1}   & $\sim$1.95        & Seen in dM and sdM                                      & M  \\
\ion{Na}{1}   & 2.207             & Strongest in dM.  Also seen in early sdM                & M  \\
CO bandheads  & 2.29              & Strongest in dM and dL.  Also seen in early sdM         & $\sim$M5.5 V, $\sim$sdM6 \\

\enddata
\tablecomments{M = M dwarf or M subdwarf, dM = M dwarf, sdM = M subdwarf, L = L dwarf or L subdwarf, dL = L dwarf, sdL = L subdwarf}

\end{deluxetable}